\documentstyle[12pt,epsf]{article}  
\newcommand{\unit}[1]{\mbox{\rm #1}}

\newcommand{\be}{\begin{equation}}  
\newcommand{\ee}{\end{equation}}

\begin{document}  
\begin{titlepage}  
\pagestyle{empty}  
\noindent
\parbox{8cm}{\sf TSL/ISV-2000-0234\\
                October 2000}
\vspace*{2cm}  
\begin{center}  
{  
\large\bf  
 Theoretical description of the total $\gamma^*\gamma^*$  
cross-section and its confrontation with the LEP data on doubly tagged  
$e^+e^-$ events  
}  
\vspace{1.1cm}\\  
 {\sc J.~Kwieci\'nski}$^a$,  
 {\sc L.~Motyka}$^{b,c}$  
\vspace{0.3cm}\\  
  
$^a${\it Department of Theoretical Physics, \\  
H.~Niewodnicza\'nski Institute of Nuclear Physics,  
Cracow, Poland}\\  
$^b${\it High Energy Physics, Uppsala University, Uppsala, Sweden} \\   
$^c${\it Institute of Physics, Jagellonian University, Cracow, Poland}  
\vspace{0.3cm}\\  
\end{center}  
\vspace{1.5cm}  

\begin{abstract}  
We perform a detailed analysis of the total inelastic cross-section  
for $\gamma^*\gamma^*$ collisions. Different contributions   
coming from the quark box diagram, reggeons, the soft and hard  
pomeron are included. The QCD pomeron contribution contains a dominant 
part of subleading effects which reduces its intercept  
and delays the onset of the asymptotic pomeron to high energies.   
Estimates of the cross-section for doubly-tagged 
$e^+e^- \to e^+e^- hadrons$ events are presented and compared
with the existing LEP data. Good  agreement between the theoretical 
results and the experimental data is found. 
We also comment on the extraction of the BFKL pomeron intercept from 
the available LEP measurements.  
\end{abstract}

\end{titlepage}  
  
\section{Introduction}  
  
The study of high energy limit of perturbative QCD is very  
interesting both theoretically and phenomenologically.   
The leading asymptotic behaviour of (semi)hard processes  
is described by the perturbative pomeron which is generated  
by the ladder diagrams with the reggeised gluon exchange along  
the chain.  The summation is carried out by the celebrated  
Balitzkij-Fadin-Kuraev-Lipatov equation \cite{BFKL,GLR}.   
Although the basic structure of the BFKL pomeron is fairly well  
understood there are still some theoretical problems which  
have to be treated with care and are not entirely solved as  
yet.  Thus  the QCD pomeron is known to  
acquire important subleading corrections \cite{BFKLNL1,BFKLNL2}.     
In particular the recently computed NLL corrections to the  
BFKL pomeron intercept are found to dominate over the leading result  
already at small values of the strong coupling constant  
$\alpha_s \sim 0.1$  that   
invalidates the perturbative expansion for this quantity.  
Thus the resumation of the perturbative series is necessary in order  
to cure this problem \cite{RESUM,RESUMCOL,KMSG,KMS}.  
Two related  approximate approaches have proved to be  
particularly useful for this purpose:  
imposing the so-called consistency constraint \cite{KMSG}, 
which follows from   
the requirement of restricting the gluon kinematics to the quasi-multiregge   
limit and the other proposed in \cite{RESUMCOL}    
which combines the leading and subleading BFKL effects with the  
renormalisation group constraints.   
Another problem is related to the fact that in the BFKL gluonic ladders   
the gluon virtualities are not restricted to the hard domain even   
if the ladder is coupled to hard objects. This is caused by the  
diffusion of the gluon transverse momenta towards the infrared region  
which is in fact enhanced   
when the running of the coupling constant in the BFKL kernel is taken   
into account \cite{GUSTAV1}. However, it was shown   
\cite{FACT} that the BFKL amplitude  can be still factorized  
into the hard and soft part(s).\\

One of the most promising measurements which can probe the BFKL pomeron   
is the determination of the total cross-section for hadronic production 
in the interaction of two virtual photons at high  
energies \cite{BRODSKY,BARTELSDR,DONAGG,DOSRGG,NIKOLGG,BARTC,PIVOV,KMGG}.  
If the virtualities of the photons are large and comparable, the process  
is fully perturbative and moreover  pure LO DGLAP evolution effects  
are suppresed due to short evolution length, leaving a  room for the genuine   
BFKL contribution to the cross-section. These conditions   
seem to be realized in the experiments measuring  
the cross-section for doubly tagged $e^+ e^-$ events at LEP1 and LEP2  
\cite{L3GG,OPAL}. The main purpose of our paper is to analyze the available  
LEP data in respect to the information about the hard pomeron that   
it contains. This of course requires  detailed treatment of   
 other competing mechanisms, like the quark box diagram contribution,   
the exchange of the soft pomeron and of the reggeons.   
Similar analysis has been performed in Ref.~\cite{DONAGG}  
where the hard pomeron term was treated  phenomenologically.    
The novel feature of our approach is the fact that the BFKL pomeron 
contribution is estimated directly from the exact (numerical) solution of  
the BFKL equation with the subleading effects generated by the  
consistency constraint taken into account.  In this way we do not  
restrict ourselves to the asymptotic form of the hard QCD pomeron 
contribution to the $\gamma^* \gamma^*$ cross-section.  It has been shown  
in Ref.~\cite{KMGG} that this asymptotic form is  
expected to be delayed to high energies.  It is therefore  
mandatory in the phenomenological analysis of  
experimental data to disentangle  the threshold effects  
from asymptotic power-like increase of the cross-sections with  
increasing energy.\\  
  
The content of our paper is as follows: in the next Section we  
recall the basic formulas connecting the process $e^+ ~ e^-  
\rightarrow  e^+ ~ e^- ~ + ~ hadrons$ with the cross-section  
for hadron production in $\gamma^* \gamma^*$ collisions, in Sec.  
3 we discuss the ``background'' to the QCD pomeron contribution  
to the total $\gamma^* \gamma^*$ cross-section,  i.e. the quark  
parton model, soft pomeron and reggeon contributions while the  
QCD pomeron contribution is discussed in Sec. 4.  In Sec. 5 we  
present comparison of our predictions with experimental data  
from LEP and in Sec. 6 we  summarise our results.

\section{Doubly tagged events}  
  
In the equivalent photon approximation the differential cross-section of   
the process $e^+e^- \to e^+e^- + hadrons$    
(averaged over the angle $\phi$ between the lepton scattering planes   
in the frame in which the virtual photons are aligned along the $z$ axis)   
is given by the following formula~\cite{BRODSKY}:  
$$  
{Q_1^2 Q_2^2 d\sigma \over dy_1 dy_2 dQ_1^2 dQ_2^2} =  
\left({\alpha\over 2 \pi}\right)^2  
[P^{(T)}_{\gamma/e^+}(y_1)P^{(T)}_{\gamma/e^-}(y_2)  
\sigma^{TT}_{\gamma^*\gamma^*}(Q_1^2, Q_2^2,W^2)+  
$$  
$$  
P^{(T)}_{\gamma/e^+}(y_1)P^{(L)}_{\gamma/e^-}(y_2)  
\sigma^{TL}_{\gamma^* \gamma^*}(Q_1^2,Q_2^2,W^2)+  
P^{(L)}_{\gamma/e^+}(y_1)P^{(T)}_{\gamma/e^-}(y_2)  
\sigma^{LT}_{\gamma^* \gamma^*}(Q_1^2,Q_2^2,W^2)+  
$$  
\begin{equation}  
P^{(L)}_{\gamma/e^+}(y_1)P^{(L)}_{\gamma/e^-}(y_2)  
\sigma^{LL}_{\gamma^* \gamma^*}(Q_1^2, Q_2^2,W^2)]  
\label{conv}  
\end{equation}  
where  
\begin{equation}  
P^{(T)}_{\gamma/e}(y) = {1 + (1-y)^2\over y}  
\label{pt}  
\end{equation}  
\begin{equation}  
P^{(L)}_{\gamma/e}(y) = 2{1-y\over y}  
\label{pl}  
\end{equation}  
In Eq.~(\ref{conv}) $y_1$ and $y_2$ are the longitudinal momentum   
fractions of the parent leptons carried by virtual photons,    
$Q_i^2 = -q_i^2$ ($i=1,2$) where $q_{1,2}$ denote the four momenta of the  
virtual photons and $W^2$ is the total CM energy squared of the  
two (virtual) photon system, i.e. $W^2=(q_1+q_2)^2$. The cross-sections  
$\sigma^{ij}_{\gamma^* \gamma^*}(Q_1^2,Q_2^2,W^2)$ are the total   
cross-sections of the process $\gamma^* \gamma^* \rightarrow hadrons$    
and the indices $i,j=T,L$ denote the  polarization of the virtual photons.    
The functions $P^{(T)}_{\gamma/e}(y)$ and $P^{(L)}_{\gamma/e}(y)$ are the   
transverse and longitudinal photon flux factors.  
  
The conditions provided by LEP detectors offer the opportunity   
to measure both the scattered electrons at the small angle $\theta$   
falling in the range of about 30 to 70~mrad. This corresponds to  
the virtualities of colliding photons:   
$Q_i^2 = 4  (1-y_i) E_{beam} ^2 \tan^2 (\theta_i /2) $.  
Since, typically $y_i \ll 1$ in such measurements the region of   
$Q^2$ equal to a few GeV$^2$ is probed at LEP1 and  above   
10~GeV$^2$ at LEP2.    
Due to rather low statistics for the doubly tagged events one focuses  
usually on a more inclusive quantity then (\ref{conv}), namely on  
\begin{equation}  
{d \bar \sigma \over dY} =   
\int dQ_1 ^2\, dQ_2 ^2 \, dy_1 \, dy_2   
{d\sigma \over dQ_1 ^2\, dQ_2 ^2 \, dy_1 \, dy_2}   
C(Q_1 ^2, Q_2 ^2, y_1 ,y_2)   
\delta \left(Y - \log\left( {y_1 y_2 s \over Q_1 Q_2} \right)\right),  
\label{barsig}  
\end{equation}  
where the function $C(Q_1 ^2, Q_2 ^2, y_1 ,y_2)$ denotes the   
experimental cuts.

\section{The background processes}  
  
In the high $Y$ and high $Q_i^2$ limit the dominant contribution   
to $\sigma^{ij} _{\gamma^* \gamma^*}$ comes from the hard pomeron  
exchange. However for non-asymptotic energies and virtualities  
the contribution of other mechanisms has to be included.  
So, when $Y$ is not large enough the quark box diagram   
contribution is important  for all virtualities.   
It becomes small for high values of $W$  decreasing as   $1/W^2$ (modulo 
logarithmic effects)  
in this limit.   
On the other hand, at low $Q^2$ non-perturbative phenomena  
i.e. the soft pomeron and for not too large values of $W$ also  
the reggeon exchange are important. We shall  analyze   
these components of the cross-section in more detail.

\subsection{The quark box contribution}  
  
The quark box (or QPM) contribution to the total $\gamma^* \gamma^*$   
cross-sections can at the leading order be calculated exactly  
and the result is given for instance in Ref.~\cite{BUDNEV}.   
The impact of the QCD corrections on the result  
has been shown to be small \cite{HillRoss,DelDuca}.   
There appears some uncertainty due to the choice of the quark masses.  
To be precise, the quark mass enters the result in two ways -- through  
the virtual quark propagators, where it would be suitable to use  
the running quark mass at the large scale $W^2$ and through the  
wave function and kinematics of the ``on-mass-shell'' produced quarks,   
where the pole mass should rather be used.   
It is not clear how to take into account both  requirements   
simultaneously so usually one set of masses is used everywhere.   
The discrepancy for the two possible choices is however small  
for photons with  virtualities above 1~GeV$^2$.   
We have chosen $m_u = m_d = m_s = 0$ and $m_c = 1.2$~GeV.   
The $b$ quark contribution is negligible due to kinematical  
effects and the low electric charge.

\subsection{The soft pomeron}  
  
In our approach the soft pomeron represents the contribution    
to the high energy amplitude coming from the exchange of gluons  
with their virtualities being in the non-perturbative domain. Since for such  
configurations the perturbative formalism does not apply  
we parametrize the corresponding contribution   using the Regge  
model. Thus we employ the Donnachie-Landshoff parameterization of the soft   
pomeron contribution to the $\gamma^* (Q^2) p$ cross-section, and   
using the Gribov factorisation hypothesis we find that the soft   
pomeron exchange gives  the following component of the   
$\gamma^*(Q_1^2) \gamma^*(Q_2 ^2)$ cross-section:  
\begin{equation}  
\sigma^{SP} _{\gamma^*\gamma^*}(Q_1^2,Q_2^2,W^2) =  
{\sigma^{SP}_{\gamma^* p}(Q_1^2,W^2)  
 \sigma^{SP}_{\gamma^* p}(Q_2^2,W^2)\over \sigma_{pp}^{SP}(W^2)}  
\label{sp}  
\end{equation}  
where $\sigma^{SP} _{\gamma^*p}$ and $\sigma_{pp}^{SP}$  
denote the soft pomeron contributions to the $\gamma^*p$ and  
$pp$ total cross-sections respectively.  
Assuming  that the soft pomeron contribution to $\gamma^* p$  
total cross-section should exhibit  
Bjorken scaling at large $Q^2$ we find:  
\begin{equation}  
\sigma^{SP}_{\gamma^*p}(Q^2,W^2) \sim {1\over Q^2}  
\left({W^2 \over Q^2}\right)^{\epsilon}  
\label{spgstp}  
\end{equation}  
where $\epsilon = \alpha_{SP}-1$ with $\alpha_{SP}$ denoting the  
soft pomeron intercept ($\alpha_{SP} \approx 1.08$).   We also have:  
\begin{equation}  
\sigma_{pp}^{SP}(W^2) \sim \left({W^2\over W_0^2}\right)^{\epsilon}  
\label{sppp}  
\end{equation}  
with $W_0 = 1$~GeV.  
From  equations (\ref{sp}, \ref{spgstp})  
and (\ref{sppp})we get:  
\begin{equation}  
\sigma_{\gamma^*\gamma^*}^{SP}(Q_1^2,Q_2^2,W^2) \sim \left({1\over Q_1^2  
Q_2^2}\right)^{1+\epsilon/2}  
\left({W^2\over Q_1Q_2}\right)^{\epsilon}  
\label{spscale}  
\end{equation}  
The characteristic feature of the soft pomeron contribution  
to the total $\gamma^*\gamma^*$ cross-section  
is its  rapid decrease  with increasing virtualities.   
It follows from equation (\ref{spscale}) that  
for large characteristic  
scale $Q^2$ for both  photons ($Q_1^2 \sim Q_2^2 \sim Q^2$),  
$\sigma^{SP}(Q^2,Q^2,W^2)$ decreases as  
$1/Q^{4+2\epsilon} \sim 1/Q^4$ (for fixed  $W^2/Q^2$)  
in contrast to the perturbative  QCD pomeron   contribution  which has  
only the $1/Q^2$ behaviour (modulo logarithmic modifications).    
The soft  pomeron contribution  should therefore be negligible at  
large $Q^2$. \\

For the numerical calculations we have used the following   
Donnachie-Landshoff parameterizations  
of $\sigma^{SP}_{\gamma^*p}(Q^2,W^2)$ and   
$\sigma_{pp}^{SP}(W^2)$ \cite{DLpp,DLgp}:  
\be  
\sigma^{SP}_{\gamma^*p}(Q^2,W^2) =  
{4\pi^2 \alpha_{em} \over a_1 + Q^2} A_1   
\left({W^2 \over  a_1 + Q^2} \right)^{\epsilon}   
\label{sgp}  
\ee  
and   
\be  
\sigma_{pp}^{SP}(W^2) = \sigma_0 \left( {W^2 \over W_0 ^2} \right)^{\epsilon}  
\ee  
where $A_1 = 0.324$, $a_1 = 0.562\unit{ GeV}^2$, $\epsilon = 0.0808$,  
$W_0 = 1$~GeV, $\sigma_0 = 21.7$~mb.  
  
\subsection{The reggeons}  
  
Exchange of the reggeons (e.g. $a_0$) is another non-perturbative   
phenomenon which can be described in terms of  an isolated Regge pole.  
The  phenomenological analysis of the total hadronic and photoproduction  
cross-sections (as well as of the  $\gamma^* p$ total cross-section)  
shows that it is characterized by the Regge intercept close to 1/2,  
yielding therefore the $\gamma^* p$ cross-sections behaving approximately  
like $1/Q^2 (W^2/Q^2)^{-0.5}$.    
Its contribution to the $\gamma^* \gamma^*$ cross-sections may be   
obtained from the  Regge pole contribution to the  $\gamma^* p$,  $pp$ and  
$p \bar p$ cross-sections   
using the Gribov factorization formula analogous to equation (\ref{sp}).  
It has to be remembered however that only the $C$-even reggeons contribute to  
the total $\gamma^* \gamma^*$ cross-section, thus we should average  
over the reggeon contribution to  $pp$ and $p\bar{p}$ cross-section   
in order to find the proton effective coupling to the  
relevant reggeons. For fixed $W$,   
the reggeon contribution to the $\gamma^* \gamma^*$ total  
cross-section  has similar dependence on the photon virtuality   
as the perturbative QCD contribution.   
The reggeon part of the cross-section decreases with  
increasing energy approximately like $1/W$.  
We have used the following formula \cite{DONAGG} for   
estimating this component:  
\be  
\sigma_{\gamma^*(Q_1^2)\gamma^*(Q_2^2)}^{R}=   
4\pi^2\alpha_{em} ^2 { A_2 \over a_2}  
\left[ {a_2^2 \over (a_2 + Q_1 ^2)(a_2+ Q_2 ^2)} \right]^{1-\eta}  
\left( {W^2 \over a_2} \right)^{-\eta}   
\ee  
with  
$A_2 = 0.38$, $a_2 = 0.3\;{\rm GeV}^2$ and $\eta = 0.45$.   
  
\section{The QCD pomeron}  
  
The hard QCD pomeron is represented by the resummed series of   
perturbative gluonic ladders which in  the leading logarithmic  
approximation is described by the BFKL equation.   
In order to take into account (in an approximate way) the non-leading   
corrections to  the BFKL kernel we use the running coupling constant along   
the gluonic ladder and the consistency constraint (CC)  
which follows from the assumption that the virtuality  
of the  gluons exchanged along the ladder is dominated by  their transverse  
momenta squared \cite{KMSG}.  
The consistency constraint was shown to introduce at the NLL   
approximation a correction to the pomeron intercept saturating  
about 70\% of the exact NLL result and  the collinear limit of the kernel  
with this constraint is consistent with the requirements of the  
renormalisation group. At least a part of the remaininig correction  
may be atributed to the running of the coupling constant.   
Therefore the LO BFKL equation with CC and running coupling constant  
may be thought of as a simplified model  
for providing the resummation of leading and subleading BFKL effects.   
Certainly, the BFKL equation constructed in the framework of perturbative  
QCD cannot hold when the gluons become too soft.  In order to eliminate  
the contribution from the non-perturbative region  
we impose a cut-off on the virtualities $k^2$ of gluons propagating  
along the ladder: $k^2>k_0 ^2 = 1 \;{\rm GeV}^2 $. This cut-off   
reflects the fact that a colour-charge cannot propagate freely in   
the QCD vacuum and its progator has a finite correlation length.  
The contribution from the non-perturbative region is treated  
phenomenologically,  
i.e. it is assummed to give the separate soft pomeron component of the  
cross-section which was discussed in the previous section.     
  
The consistency constraint restricts the available phase-space   
for the emissions of real gluons --- thus suppressing the   
radiation. The most important consequence is reduction  
of the pomeron intercept. However it also  
interlocks the longitudinal  
and transverse components of the gluon momenta. This makes it   
possible to define an energy scale for the onset  
of the Regge regime, since the distribution of the transverse gluon  
momenta has a natural characteristic scale.\\

In order to estimate the contribution of the  
QCD pomeron it is convenient, following  
our previous treatment in Ref.~\cite{KMGG}, to introduce the unintegrated  
gluon distributions $\Phi_i(k^2,Q^2,x_g)$ in the virtual photon  
of virtuality $Q^2$ where $k^2$ and $x_g$ denote the gluon transverse  
momentum squared and the longitudinal momentum fraction of the  
parent  virtual photon carried by the gluon respectively.  The index $i$  
corresponds to the transverse or longitudinal polarisation of  
the virtual photon.  The unintegrated gluon distribution  
$\Phi_i(k^2,Q^2,x_g)$ satisfies the (modified) BFKL equation which reads:  
$$  
\Phi_i(k^2,Q^2,x_g)=\Phi^0_i(k^2,Q^2,x_g)+\Phi^S(k^2,Q^2,x_g)\delta_{iT}+  
{3\alpha_s(k^2)\over \pi} k^2\int_{x_g}^1 {dx^{\prime}\over x^{\prime}}  
\int_{k_0^2}^{\infty} {dk^{\prime 2} \over k^{\prime 2}}  
$$  
\begin{equation}  
\left [ {\Phi_i(k^{\prime 2},Q^2,x^{\prime})\Theta  
\left(k^2{ x^{\prime}\over x_g}  
-k^{\prime 2}\right) - \Phi_i(k^{ 2},Q^2,x^{\prime})  
\over |k^{\prime 2} - k^{ 2}|} + {\Phi_i(k^{ 2},Q^2,x^{\prime})\over  
\sqrt{4 k^{\prime 4} +  k^{4}}}\right]  
\label{bfklcc}  
\end{equation}  
where the  function  
$\Theta \left(k^2{ x^{\prime}\over x_g} -k^{\prime 2}\right)$  
reflects the consistency constraint.   
The inhomogeneous term $\Phi^0_i(k^2,Q^2,x_g)$ corresponds to  
the quark box and crossed box contribution  
to the unintegrated gluon distribution in the photon  
and  the term $\Phi^S(k^2,Q^2,x_g)\delta_{iT}$ corresponds to  
the soft pomeron contribution to this distribution.    
The detailed definition of those two functions is given 
in Ref.~\cite{KMGG}.\\  
  
The QCD pomeron contribution to the $\gamma^* \gamma^*$ total  
cross-sections  $\sigma^{ij}_{\gamma^* \gamma^*}(Q_1^2, Q_2^2,W^2)$ is  
given by the following formula:  
$$  
\sigma^{ij}_{\gamma^* \gamma^*}(Q_1^2,Q_2^2,W^2) =  
$$  
\begin{equation}  
{1\over 2 \pi}\sum_q\int_{k_0^2}^{k_{max}^2(Q_2^2,x)} {dk^2\over k^4}  
\int _{\xi_{min}(k^2,Q_2^2)}^{1/x} d\xi  
 G^{0j}_q(k^2,Q_2^2,\xi)  
 \Phi_i(k^2,Q_1^2,x\xi)  
\label{csx}  
\end{equation}  
where  
\begin{equation}  
k_{max}^2(Q_2^2,x)=-4m_q^2+Q_2^2 \left( {1\over x} - 1 \right)  
\label{kmax}  
\end{equation}  
\begin{equation}  
\xi_{min}(k^2,Q_2^2)=1+{k^2+4m_q^2\over Q_2^2}  
\label{ximin}  
\end{equation}  
and  
\begin{equation}  
x={Q_2^2\over 2q_1q_2}  
\label{x}  
\end{equation}  
The impact factors $G^{0j}_q(k^2,Q_2^2,\xi)$ are defined in  
Ref.~\cite{KMGG}. The increase of the function  
$\Phi_i (k^2, Q_1^2,x_g)$ with decreasing $x_g$ which follows from  
the BFKL equation, generates increase of the cross-section  
$\sigma_{\gamma^*\gamma^*} (Q_1^2, Q_2^2,W^2)$  
with increasing $W^2$. This is implied by the fact that  
$x_g \sim x$ (cf. Eq.~(\ref{csx})).

Let us recall that in the conventional leading $\log(1/x)$ approximation   
we should set $\Phi^0_i(k^2,Q^2,x_g=0)$ in place of  $\Phi^0_i(k^2,Q^2,x_g)$  
in the BFKL equation (\ref{bfklcc}).   
It is also legitimate in this  approximation   
to set just $x$ instead of $ x \xi$ as the argument of  
$\Phi_i$ in equation (\ref{csx}) and neglect all phase space  
limitations constraining integrations over $d\xi$ and $dk^2$.   
These approximations lead to the following approximate  
expression for the $\gamma^*\gamma^*$ cross-sections:  
\begin{equation}  
\sigma^{ij}_{\gamma^* \gamma^*}(Q_1^2,Q_2^2,W^2) =  
{1\over 2 \pi}\sum_q\int_{k_0^2}^{\infty} {dk^2\over k^4}  
\Phi^0_i(k^2,Q^2,x_g=0)\bar \Phi_i(k^2,Q_1^2,x)  
\label{csxa}  
\end{equation}  
where  the function $\bar \Phi_i(k^2,Q_1^2,x)$ corresponds to the  
solution  of the BFKL equation with the inhomogeneous term  
equal to  $\Phi^0_i(k^2,Q^2,x_g=0)$ instead of $\Phi^0_i(k^2,Q^2,x_g)$.  
It should be emphasised that formula (\ref{csx}), which our  
estimate of the QCD pomeron contribution is based upon contains  
important  kinematical effects which are missing   
in its approximate asymptotic form (\ref{csxa}).   
First of all equation (\ref{csx}) includes  
the phase space effects generated by the kinematical  
limits constraining   integrations over $d\xi$ and $dk^2$.   
Moreover we do also keep $x_g$ dependence  in the impact factors  
$\Phi^0_i(k^2,Q^2,x_g)$ defining the inhomogeneous term of the BFKL  
equation. The variable  $x_g$ is limited from below by $x$.    
It provides a kinematical lower bound for the longitudinal momentum  
fraction $z_q$ of the  quark emitting the gluon in the quark  
box diagram defining the impact factors $ \Phi^0_i(k^2,Q^2,x_g)$,  
i.e. $z_q > x_g > x$ \cite{KMGG}.  By keeping  
the $x_g$ dependence of the impact factors $ \Phi^0_i(k^2,Q^2,x_g)$  
we  generate corrections to their asymptotic  
form in the limit $x_g \to 0$   
which are subleading (have an additional factor of $x_g$) but  
non-negligible even for low values of $x_g$.  
 
\begin{figure}[hbpt] 
\begin{center} 
\leavevmode 
\epsfxsize = 13.0cm 
\epsfysize = 13.0cm 
\epsfbox[20 255 550 790]{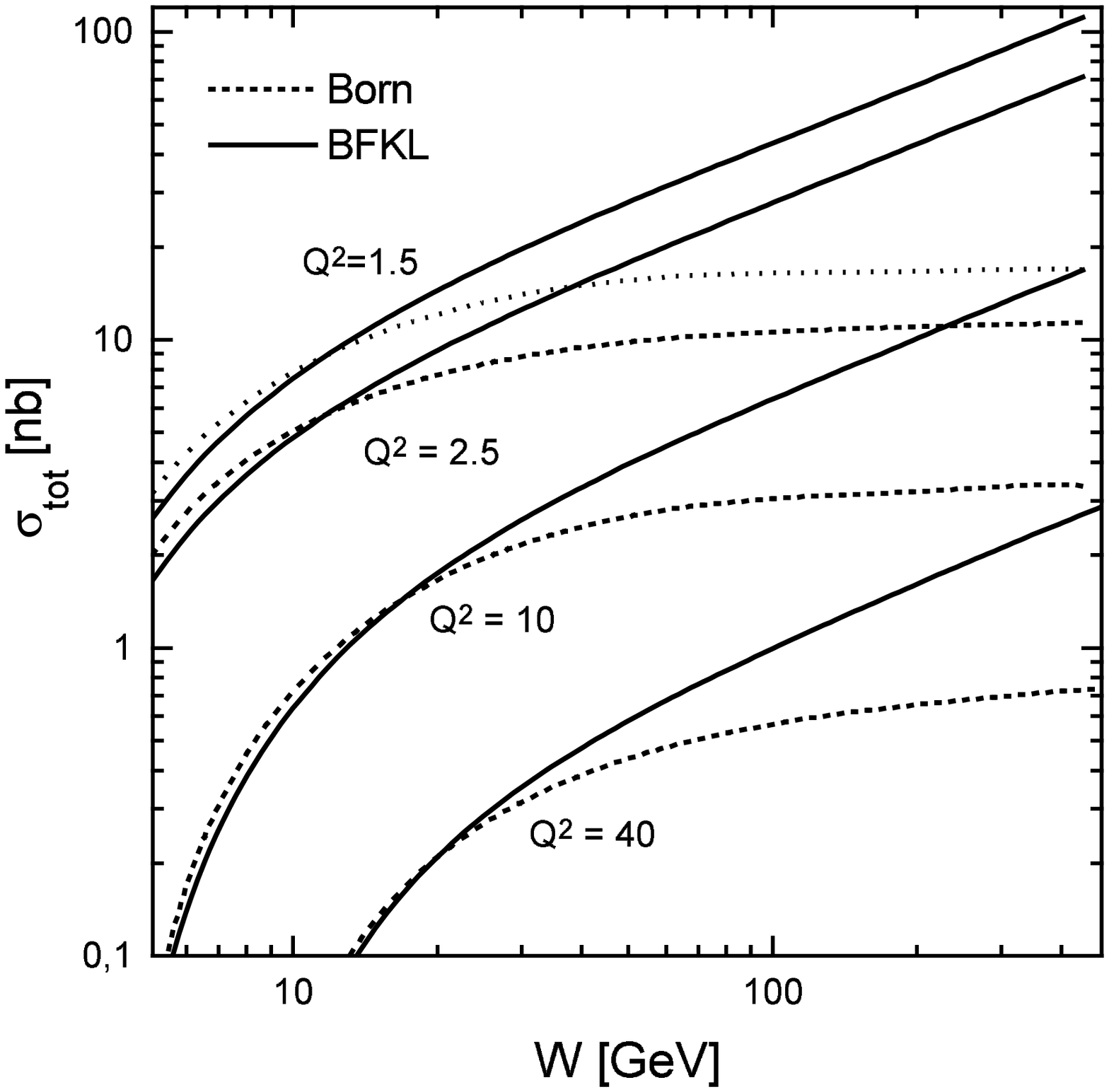} \\ 
\end{center} 
\caption{ \small Energy dependence of the cross-section 
$\sigma^{TT} _{\gamma^* \gamma^*}(Q_1 ^2, Q_2 ^2, W^2)$ 
for the process $\gamma^*(Q_1 ^2) \gamma^*(Q_2^2)  \to hadrons$ 
for various choices of photon virtualitities $Q^2 = Q_1 ^2 = Q_2 ^2$: 
Comparison of the complete contribution of the perturbative QCD pomeron   
to the cross-section $\sigma^{TT} _{\gamma^* \gamma^*}(Q ^2, Q ^2, W^2)$ 
(continous line) with its Born term component corresponding to the two gluon 
exchange mechanism (dotted line). The figure from Ref.~\cite{KMGG}.  } 
\label{fullborn} 
\end{figure}

These effects, i.e. the phase-space limitations in equation (\ref{csx})  
and the $x_g$ dependence of the impact factors  $\Phi^0_i(k^2,Q^2,x_g)$  
delay the onset of the asymptotic  (power-like) behavior of the total  
cross-sections.   
They also introduce an additional energy dependence of the  
cross-sections in the low~$W$  region which is essentially of kinematical  
origin. In particular, they are entirely responsible for generating  
sub-asymptotic energy dependence of the Born term (i.e. given by the  
two gluon exchange) which in the high energy limit gives the constant  
cross-section (see Fig.~\ref{fullborn}).  
Finally, it is worthwile to note  that such threshold effects may be  
misidentified as the asymptotic, power-like rise of the cross-section giving   
too high value of the pomeron intercept.    \\

It follows from our previous analysis \cite{KMGG} of the QCD pomeron   
contribution to $\gamma^*\gamma^*$ total cross-section(s) that the 
resulting pomeron intercept equals  
about 0.35 whereas the asymptotic, power-like behaviour is  
reached roughly when $W^2 > 20 \;Q_1 Q_2$ (see Fig.\ref{fullborn}).   
In terms of the commonly used variable  $Y$ this means that we should 
expect the power-like rise of the cross-section   
to start only  at $Y>3$ and giving a substantial effect for $Y>5$.  
Such effect seems to appear in the recent L3 data \cite{L3GG}.  
  
The definition of the impact factors is straightforward in the leading  
order approximation of perturbative   
QCD in which they are given by the quark box diagrams contribution.   
In this approximation the impact factors are proportional to  
$\Phi^0_i(k^2,Q^2,x_g=0)$. However the leading order expressions may be 
affected by the possibly important subleading corrections which are not 
known yet 
\footnote{In our framework we take into account a part of the subleading  
corrections which are of the kinematical origin.}$^,$ 
\footnote{The calculation of  
the NLO corrections to the impact factors is in progress \cite{IFNLO}.}.   
Thus the non-leading corrections introduce some uncertainty into our result.  
Besides that the value of the   
scale for the running strong coupling constant describing the interaction   
between quarks and gluons is ambigous. In fact both these problems are 
related and an improvement here can only be achieved when the non-leading  
corrections to the impact-factors become known.  
A natural choice for the scale $\mu^2$ of the running   
coupling constant $\alpha_s(\mu^2)$   would be the virtuality   
of the interacting gluon $k^2$ ($k^2>0$), possibly combined with the   
relevant quark mass $m_q$ squared: $\mu^2 =  k^2 + m_q ^2$.  
However, because there is some freedom left here one may use also  
an alternative choice of $\mu^2$, e.g. $\mu^2 =  ( k^2 + m_q ^2) / 4$  
in order to check the uncertainty of the prediction and possibly  
find an experimental hint on the proper choice of the scale.  
These uncertainties do only affect normalisation of the cross-sections  
leaving unchanged their energy dependence.  
We have confronted the predictions obtained with the use of   
both  scenarios with the experimental data from LEP.   
The lower scale scenario yields the $\gamma^*\gamma^*$ cross-section 
two times  
bigger than that obtained with the standard choice of $\mu^2 = k^2 + m_q ^2$  
but  still slightly below the data. Thus it is clear   
that at present the LEP data favour taking the low scale in the impact   
factors. Of course this conclusion may be altered if the   
NLO corrections to impact factors turn out to be positive and large.

\section{Comparison with the data}  
  
\subsection{Do we see the pomeron?}  
\begin{figure}[hbpt] 
\parbox{17cm}{ 
\begin{center} 
\leavevmode 
\noindent 
\hspace{-5mm} 
\epsfxsize = 7.5cm 
\epsfysize = 7.5cm 
\epsfbox{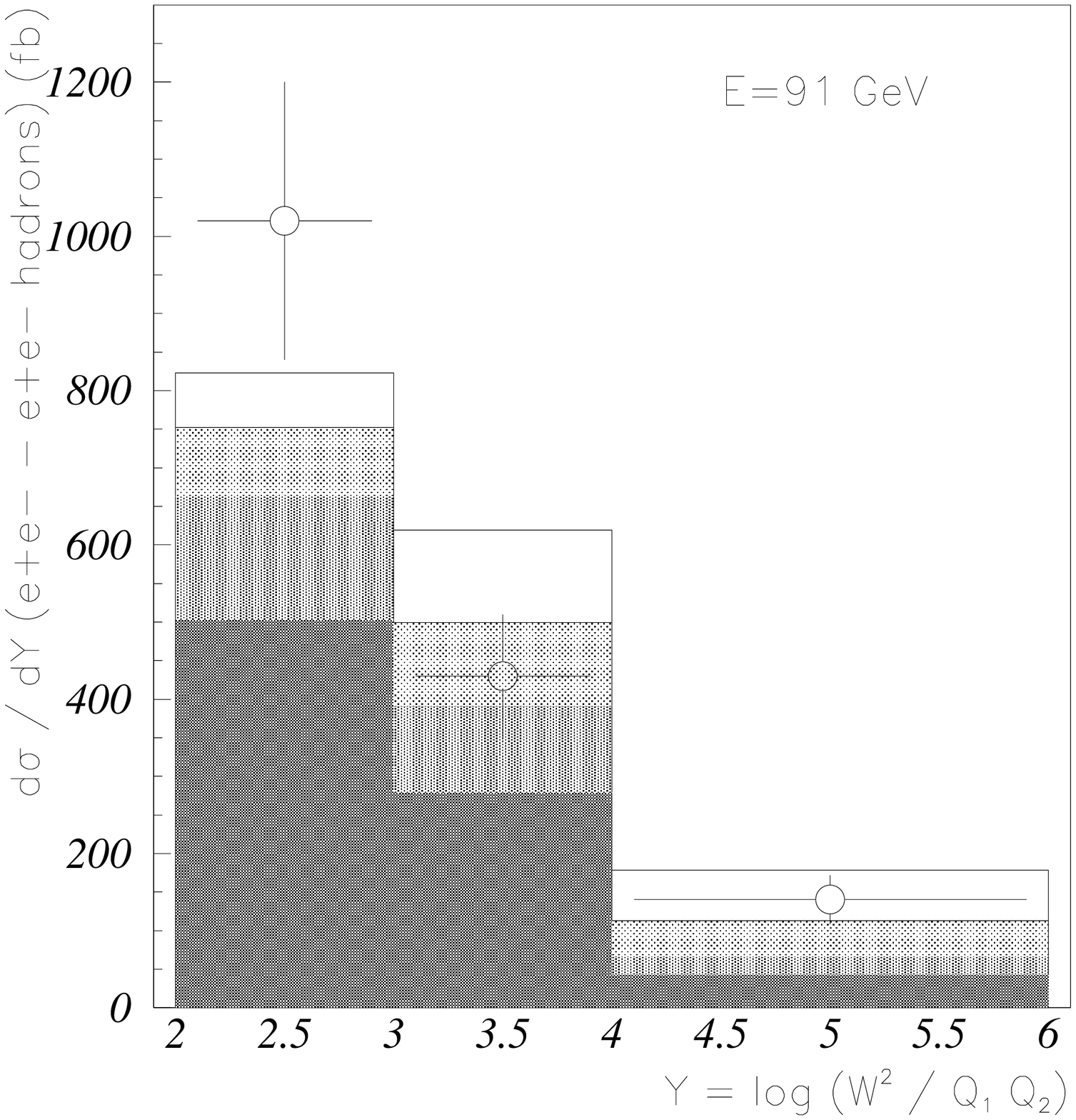} \hspace{5mm} 
\epsfxsize = 7.5cm 
\epsfysize = 7.5cm 
\epsfbox{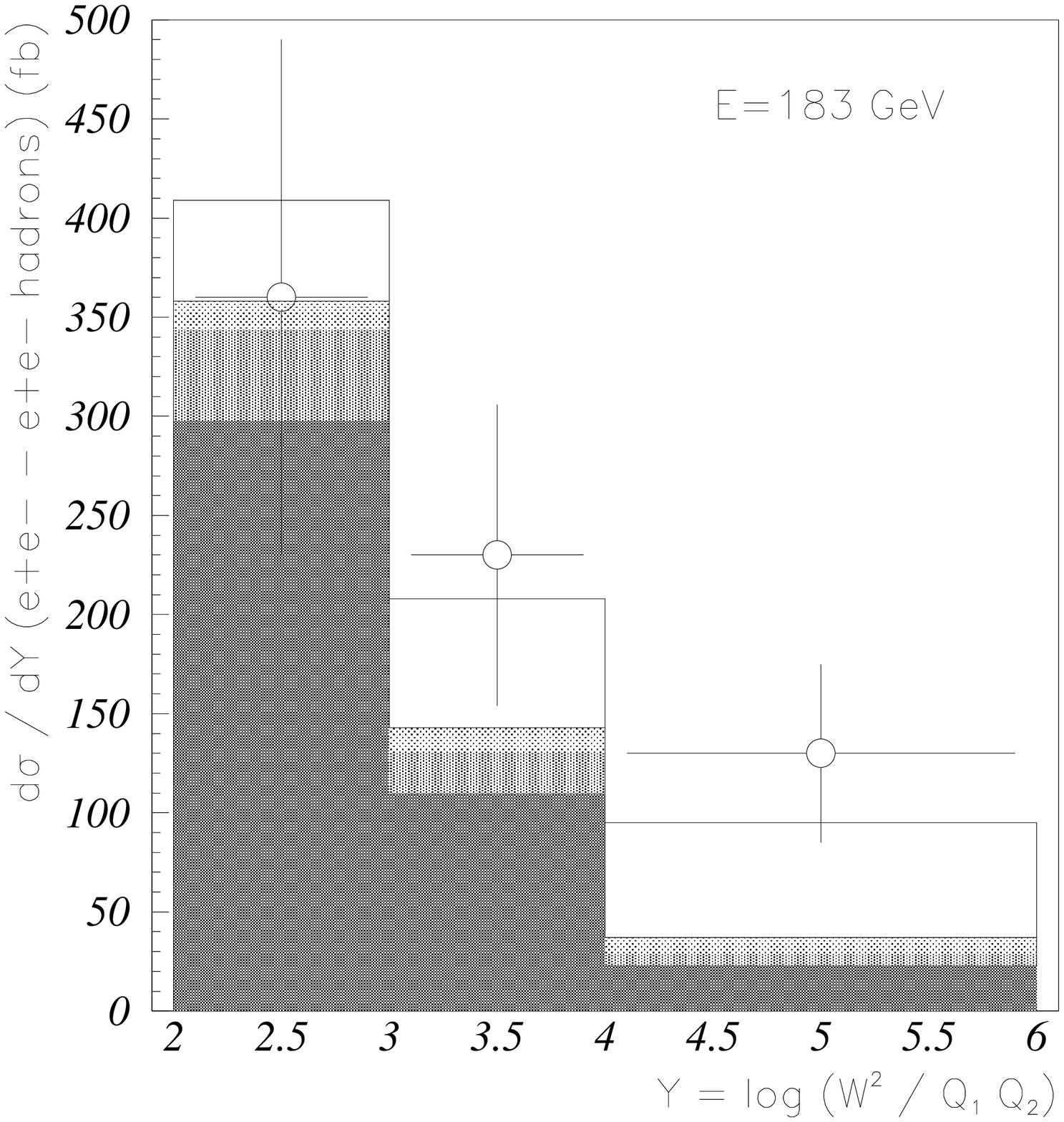}  \\[10mm] 
\hspace{-5mm} 
\epsfxsize = 7.5cm 
\epsfysize = 7.5cm 
\epsfbox{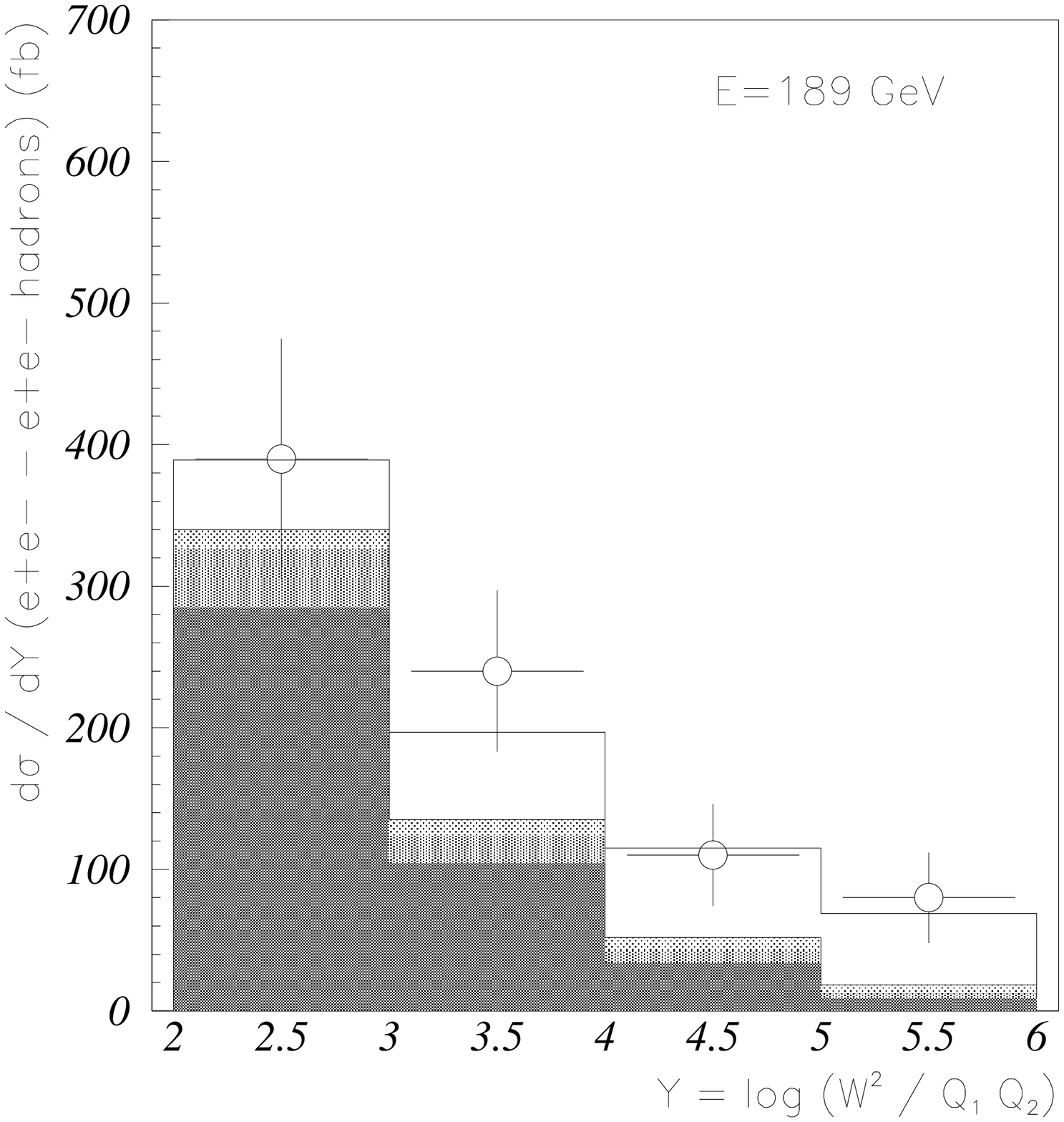} \hspace{5mm} 
\epsfxsize = 7.5cm 
\epsfysize = 7.5cm 
\epsfbox{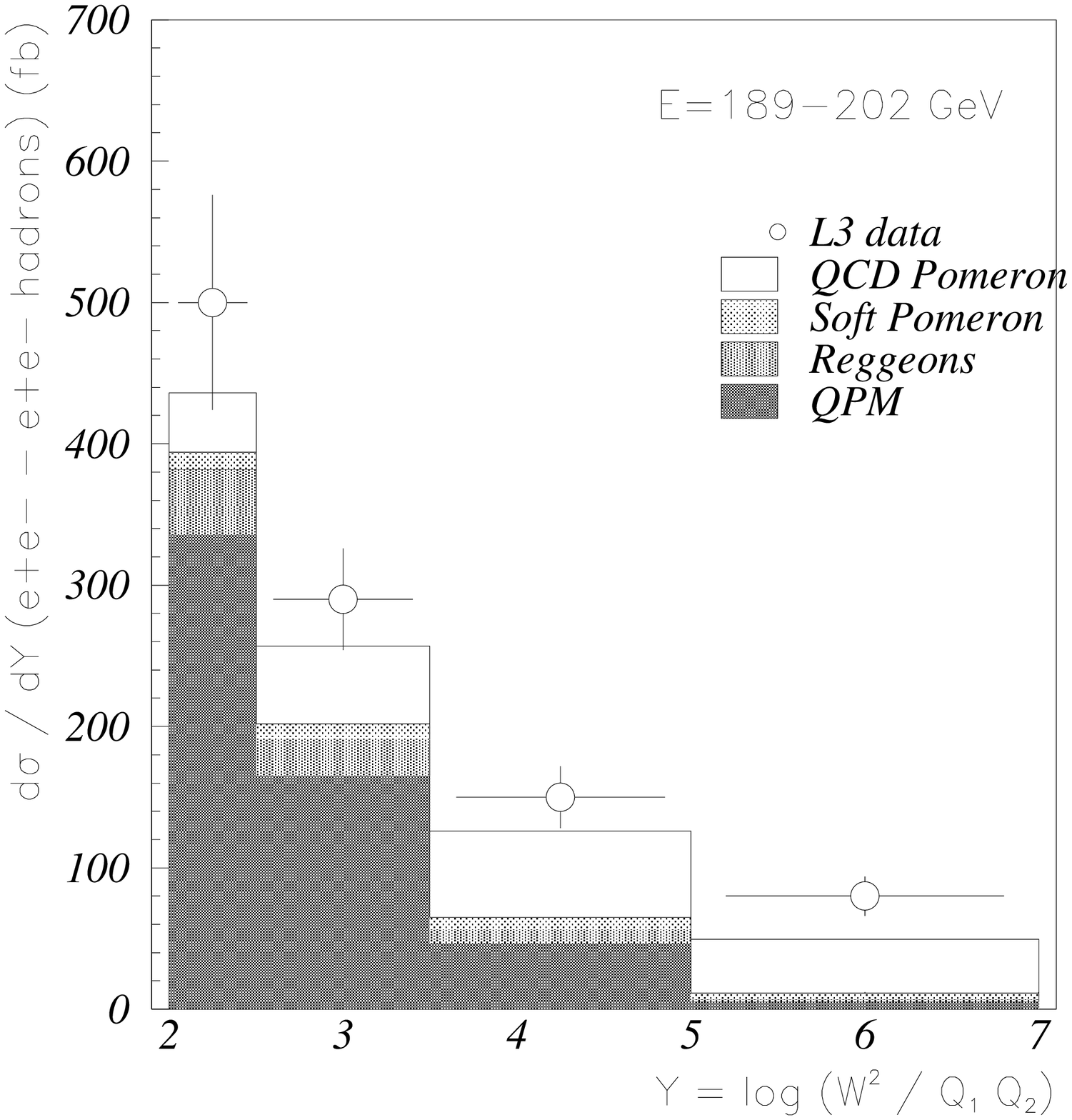}  
\end{center}} 
\caption{ \small  
Comparison of the L3 data \cite{L3GG}
on the differential cross-section for doubly  
tagged events $d\sigma(e^+e^-\to e^+e^-\;+\; hadrons)/dY$ 
with our predictions plotted as function of $Y$ for different  
$e^+e^-$ collision energies, corresponding to the measurements by the 
L3 collaboration. Four different mechanisms contributing to 
$d\sigma/dY$ are described in the text. 
} 
\label{L3data} 
\end{figure}

\begin{figure}[hbpt] 
\begin{center} 
\leavevmode 
\epsfxsize = 13cm 
\epsfysize = 13cm 
\epsfbox{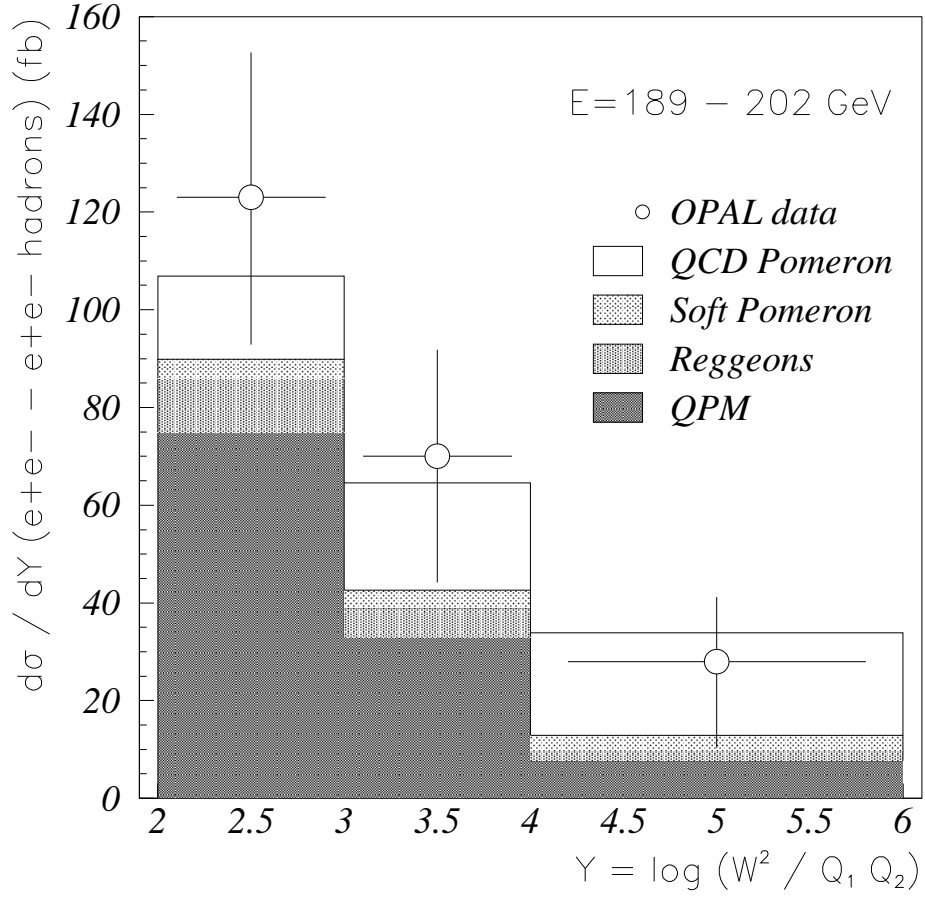} \\ 
\end{center} 
\caption{ \small  
Comparison of the OPAL preliminary data \cite{OPAL} 
on the differential cross-section for doubly  
tagged events $d\sigma(e^+e^-\to e^+e^-\;+\; hadrons)/dY$ 
with our predictions plotted as function of $Y$ for  
the $e^+e^-$ collision energies between 189 and 202 GeV. 
} 
\label{OPALdata} 
\end{figure}

L3 and OPAL collaborations have collected a sample of data for double  
tagged events both at the $Z^0$ peak and for the $e^+e^-$ between 183 and  
202 GeV \cite{L3GG,OPAL}. In Fig.~\ref{L3data} and Fig.~\ref{OPALdata} we  
give the comparison between the theoretical results obtained in the  
framework of our model and the data.   
Our theoretical predictions include  all components of the 
$\gamma^*\gamma^*$ cross-section  discussed  in previous sections  
and are obtained using formula (\ref{barsig}) with  
L3 and OPAL cuts and binning respectively. It can be seen that the model  
reproduces  the experimental results rather well although we tend to  
overestimate the L3 data from LEP1 and underestimate those from LEP2.   
We point out that the hard pomeron  dominates the cross-section  
for $Y> 3.5 - 4$  so this region of $Y$ is particularly interesting.  
Moreover, in the LEP1 data the non-perturbative contributions  
coming from the soft pomeron and the reggeons are still of   
considerable importance whereas at LEP2 energies they become  
of little relevance. This is fortunate because these   
non-perturbatively driven components of the cross-section are known   
with a rather limited accuracy due to errors of fits and some   
theoretical ambiguities  (in particular the interplay between the soft   
and hard pomeron). On the other hand the quark-box contribution is  
known very accurately so the uncertainty which is introduced  
when subtracting this component from the total cross-section  
is small. Therefore the LEP2 data, especially those for $Y> 3.5 - 4$  
carry the most precise information about the QCD pomeron.   
It is also in the large $Y$ region where we expect the significant  
BFKL enhancement of the cross-section.

Let us also quote comparisons between the data and the PHOJET  
Monte-Carlo program \cite{PHOJET} presented by L3 and OPAL.  
PHOJET contains all the components of the $\gamma^*\gamma^*$  
cross-section but  the ``hard pomeron'' part is decribed there in the  
framework of DGLAP evolution only while the BFKL-type effects are neglected.  
Since the virtualities of both  photons are comparable this  
essentially corresponds to the two gluon exchange process.  
Therefore, if our conclusions are correct,  it should work rather well   
up to $Y=4$. Indeed it does but already for $4<Y<6$ (at $\sqrt{s}=183$~GeV)   
the data published by both the LEP experiments are slightly underestimated  
and for the L3 measurement at $5<Y<7$ (at $\sqrt{s}>189$~GeV)  
the PHOJET prediction of $d\sigma / dY = 30$~fb is below the data   
($d\sigma / dY_{exp} = 80\pm 10 \pm 10$~fb) by more than three standard   
deviations. Our result in this bin reads 52~fb being by   
two standard deviations below the preliminary experimental point.    
However (see Fig.~\ref{L3data}) we  describe rather well the   
energy dependence of the pomeron contribution, perhaps  
underestimating the overall normalisation of this term.       
The OPAL data have larger error bars and cover a smaller  
region of $Y$ than the L3 data. So, the tendency reported by OPAL that  
$d\sigma/ dY (Y)$, estimated with the PHOJET MC, underestimates the  
central values of the experimental points at high $Y$ is not yet 
statistically significant. However, our model reproduces the OPAL 
data better, i.e. has lower $\chi^2$, than PHOJET.  
The data from both the experiments confirm also 
our prediction that the onset of the power-like increase 
of the $\gamma^* \gamma^*$ cross-section is delayed 
to $Y>3.5$.   
  
Let us finally point out  the problem with radiative QED corrections to the  
double tagged cross-section. It has  been  noticed in Ref.~\cite{OPAL}, that   
if the $\gamma^*\gamma^*$ collision energy $W$ is obtained directly from  
a measurement of the kinematics of tagged electrons, as done by L3, the  
estimated  and actual value of $W$ may be different.  
This discrepancy is  mainly caused by the electromagnetic initial state  
radiation since the emitted ISR photons which have small enough virtuality  
are not seen by the detector. Nevertheless they can carry a non-negligible  
fraction of the lepton energy, affecting therefore the determination of $W$  
based on the measurement of leptonic four-momenta. This effect is supressed  
by a small factor $\alpha_{em}/\pi$ but after performing an integration over  
the photon virtuality giving  a large logarithm, it appears to be sizable  
\cite{OPAL}. According to our knowledge the L3 collaboration has not  
included this important correction in the analysis. This means that the 
L3 data, especially for large $Y$ may change, namely $d\sigma /dY$ 
may turn out to be smaller. The OPAL collaboration uses  
produced hadronic invariant mass to calculate $W$ and thus the OPAL data 
are not affected by such a systematic error.   
The consistency between our predictions and the OPAL data is very good.

\subsection{On the determination of the intercept}  
One of the important aims of  
the  empirical analysis of  experimental data on the  
reaction $e^+e^- \rightarrow e^+e^- + hadrons$ with tagged leptons  
is the determination of the effective QCD pomeron intercept which  
describes the high $W$ behaviour of the $\gamma^*\gamma^*$ cross-section.  
In order to extract directly this cross-section from the experimental   
results with doubly tagged events it is necessary to perform a deconvolution  
using  photon luminosity functions. Furthermore, one has to subtract   
the remaining components of the cross-section. Then the results are usually  
given  
as a function $d\sigma_{\gamma^*\gamma^*} / dY$ of the variable Y and  
the asymptotic form of the  BFKL cross-section    
\be  
{d\sigma_{\gamma^*\gamma^*} \over  dY} \sim {1\over Q_1 Q_2}  
{\exp[(\alpha_{P}-1)Y] \over \sqrt{Y}}  
\label{ABFKL}  
\ee  
is fitted with the pomeron intercept $\alpha_P$ being left as   
a free parameter.

\begin{figure}[hbpt] 
\noindent 
\begin{center} 
\leavevmode 
\epsfxsize = 9cm 
\epsfysize = 9cm 
\hspace{1mm} \epsfbox{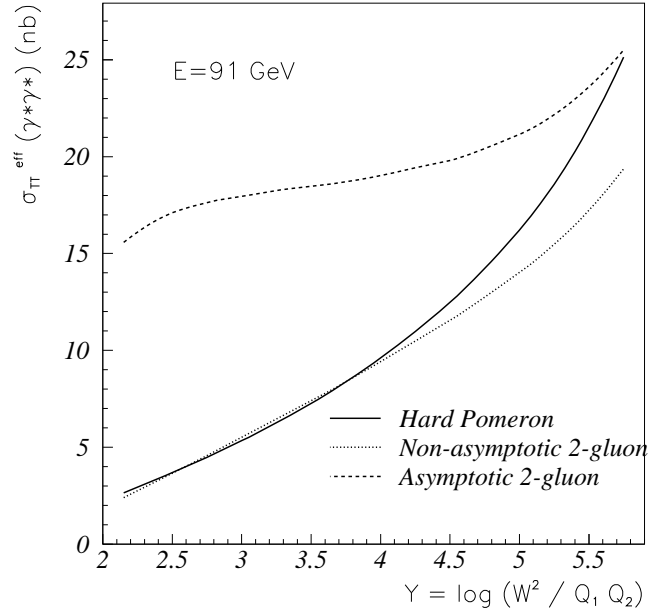} \\ 
\epsfxsize = 9cm 
\epsfysize = 9cm  
\hspace{1mm} \epsfbox{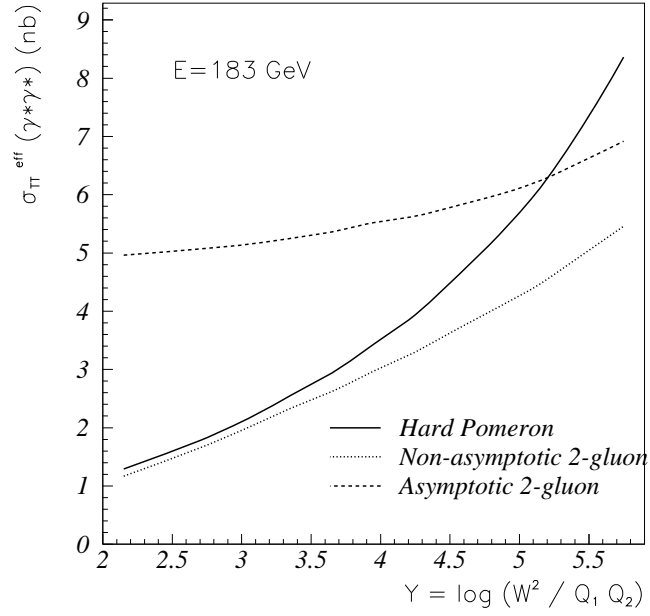} \\ 
\end{center} 
\caption{ \small  
Effective $\gamma^* \gamma^*$ cross-section defined by formula 
(\ref{barggsig}) for transverse photons obtained with L3 cuts 
for two different $e^+e^-$ energies. Three curves are shown 
corresponding to the hard pomeron (continous line), two gluon exchange 
(dotted line) and the two gluon exchange in the limit of asymptotically 
high energies (dashed line) approximations of  
$\sigma^{TT} _{\gamma^* \gamma^*}(Q_1^2,Q_2^2,W^2)$. 
} 
\label{Effective} 
\end{figure}

We would however like to point out that this method of  
determining the pomeron intercept may not be correct.   
We understand that in the fitting procedure it is usually  
assumed that the $Q_i^2$ distributions are the same in each of the  
$Y$-bins and do not affect the  fits  of $Y$ dependence. However,  
the variable $Y$ is in fact correlated  
with $Q_1 Q_2$ and we may expect the distribution of $Q_1 Q_2$ to be  
dominated  by the low photon virtualities for high $Y$ bins and  
by the high values of the virtualities for low $Y$.  
In this case we would obtain some  
enhancement of the effective $\gamma^* \gamma^*$ cross-section for high     
$Y$ due to strong $1/(Q_1 Q_2)$ dependence of the total cross-section.   
This effect may imitate the genuine increase of total cross-sections  
with increasing $W^2$. In Fig.~\ref{Effective}  
we show that such  effect indeed occurs.   
In this figure we plot the effective cross-sections  
$\bar \sigma_{\gamma^*\gamma^*} ^{TT}  (Y)$ (for the L3 cuts)   
 obtained with different assumptions concerning    
$\sigma_{\gamma^*\gamma^*} ^{TT}  (W^2,Q_1^2,Q_2^2)$.   
We used the following definition of    
$\bar \sigma_{\gamma^*\gamma^*} ^{TT}  (Y)$ (cf.~formula (\ref{barsig})):  
\be  
\bar \sigma_{\gamma^*\gamma^*} ^{TT}  (Y) =   
{ \int dQ_1 ^2 \, dQ_2 ^2 \, dy_1 \, dy_2  
       U(y_1,y_2,Q_1^2,Q_2^2;Y)   
       \sigma^{TT}_{\gamma^* \gamma^*} (W^2,Q_1^2,Q_2^2)    
\over  
\int dQ_1 ^2 \, dQ_2 ^2 \, dy_1 \, dy_2   
     U(y_1,y_2,Q_1^2,Q_2^2;Y)  }  
\label{barggsig}  
\ee  
where the weight factor $U$ reads  
\be   
U(y_1,y_2,Q_1^2,Q_2^2;Y) =    
 C(Q_1 ^2, Q_2 ^2, y_1 ,y_2)  
 { P^{(T)}_{\gamma/e}(y_1) \over Q_1 ^2}   
 { P^{(T)}_{\gamma/e}(y_2) \over Q_2 ^2}  
 \delta \left(Y - \log\left( {y_1 y_2 s \over Q_1 Q_2} \right)\right).  
\ee  
The effective $\gamma^*\gamma^*$ cross-section defined in this way   
should directly correspond to the experimental measurements  
of  $ \sigma_{\gamma^*\gamma^*} ^{TT}  (Y) $.

The modification of the $Y$ dependence can be seen at best  
when looking on the dashed curves which correspond to the  
asymptotic two-gluon contribution which gives completely flat $W$ dependence.  
The different  
magnitude of photon virtualities in different   
$Y$-bins generates the effective intercept $\bar \alpha_P \sim 1.1$ at  
LEP1 conditions and  $\bar \alpha_P \sim 1.05$ at LEP2. So, the conclusions  
based on the simple fits to  $ \bar\sigma_{\gamma^*\gamma^*}  (Y)$ would   
lead to overestimate of the intercept, whose true value was $\alpha_P =1$.   
This is  not a dramatic effect but still goes beyond the claimed accuracy  
of the determination of the intercept. A simple remedy to remove   
such problems would be to use in the fit the quantity   
$\langle Q_1 Q_2  \rangle (Y) \;  \bar\sigma_{\gamma^*\gamma^*}(Y)$   
for the pomeron mediated part of the total $\sigma_{\gamma^*\gamma^*}(Y)$,  
instead of $\bar\sigma_{\gamma^*\gamma^*}(Y)$.   
The factor $\langle Q_1 Q_2 \rangle (Y)$ stands for the   
mean values of $ Q_1 Q_2$ in the given $Y$ bin. Then the sensitivity   
to the virtualities should become  substantially reduced.

\section{Summary and conclusions}  
  
In this paper we have performed a theoretical analysis of the  
$\gamma^* \gamma^*$ total cross-section assuming the QCD pomeron  
exchange together with the contributions given by the soft pomeron,  
the non-pomeron reggeons and by the QPM term.   
The QCD pomeron contribution was  
calculated from the numerical solution of the modified BFKL equation  
in which we have included the dominant subleading effects generated  
by the consistency constraint limiting the available phase space  
(see Eq.~(\ref{bfklcc})).  We have also included phase space effects  
in the corresponding formula which conects the total $\gamma^* \gamma^*$  
cross-section(s) with the solution of the BFKL equation (see Eq.~(\ref{csx})).  
These effects delay the onset of the asymptotic QCD pomeron contribution.   
The soft pomeron and non-pomeron reggeon  
terms were estimated using Gribov factorisation.   
Our theoretical predictions have been obtained with only one adjustable  
parameter characterizing the energy scale $\mu^2$ for the running  
coupling constant in the impact factors and  were found to give a  
reasonable description  of the experimental data from LEP.   
We have found that the soft pomeron and reggeon contributions are  
important at LEP1.  They are however  
less significant at LEP2 over the entire $Y$ range.   
This is linked with the fact that the magnitudes of the virtualities  
$Q_{1,2}^2$ which are sampled at LEP2 are larger than those at LEP1.   
The QPM contribution is very important for low values of $Y$.   
The region of large values of $Y$ is dominated by the QCD pomeron  
contribution. To summarise we have found that although the  
QCD pomeron exchange mechanism  is important for the description of the LEP  
data the other contributions, i.e. QPM and the soft pomeron and non-pomeron  
reggeons are non-negligible and should be included in the analysis.   
The QCD pomeron should however give the dominant contribution at energies  
which  will be probed in future linear $e^+e^-$ colliders.  
  
\section*{Acknowledgments}  
We are indebted to Gunnar Ingelmann for critically reading the manuscript 
and valuable comments. We thank Sandy Donnachie, Mariusz Przybycie\'{n} and 
Albert De Roeck for very useful and inspiring discussions. 
LM is grateful to the Swedish Natural Science Research Council 
for the postdoctoral fellowship.
This research was partially supported  
by the EU Fourth Framework Programme `Training and Mobility of Researchers',  
Network `Quantum Chromodynamics and the Deep Structure of Elementary  
Particles', contract FMRX--CT98--0194 and  by the Polish 
Committee for Scientific Research (KBN) grant NO 2P03B 05119.

\newpage

\end{document}